\documentclass[preprint,aps,tightenlines,showpacs,nofootinbib]{revtex4}
\usepackage{latexsym}

\newcommand{\beq}{\begin{equation}}
\newcommand{\eeq}{\end{equation}}
\newcommand{\beqa}{\begin{eqnarray}}
\newcommand{\eeqa}{\end{eqnarray}}
\newcommand{\sr}{\sqrt}
\newcommand{\fr}{\frac}

\begin{document}

\preprint{hep-th/0402113, KIAS-P03018, WU-AP/158/03 }

\title{
Near-Horizon Conformal Symmetry and Black Hole Entropy in Any
Dimension}

\author{
Gungwon Kang$^{a,}$\footnote{E-mail address: gwkang@kias.re.kr},
Jun-ichirou Koga$^{b,}$\footnote{E-mail address:
koga@gravity.phys.waseda.ac.jp}, and Mu-In
Park$^{c,}$\footnote{E-mail address: muinpark@yahoo.com}}

\affiliation{$^a$School of Physics, Korea Institute for Advanced
Study,
207-43 Cheongryangri-dong, Dongdaemun-gu, Seoul 130-012, Korea \\
$^b$Advanced Research Institute for Science and Engineering,
Waseda University, Shinjuku-ku, Tokyo 169-8555, Japan  \\
$^c$Department of Physics, POSTECH, Pohang 790-784, Korea}

\begin{abstract}

Recently, Carlip proposed a derivation of the entropy of the
two-dimensional dilatonic black hole by investigating the Virasoro
algebra associated with a newly introduced near-horizon conformal
symmetry. We point out not only that the algebra of these
conformal transformations is not well defined on the horizon, but
also that the correct use of the eigenvalue of the operator $L_0$
yields vanishing entropy. It has been shown that these problems
can be resolved by choosing a different basis of the conformal
transformations which is regular even at the horizon. We also show
the generalization of Carlip's derivation to any higher
dimensional case in pure Einstein gravity. The entropy obtained is
proportional to the area of the event horizon, but it also depends
linearly on the product of the surface gravity and the parameter
length of a horizon segment in consideration. We finally point out
that this derivation of black hole entropy is quite different from
the ones proposed so far, and several features of this method and
some open issues are also discussed.

\end{abstract}


\maketitle

\newpage

\section{Introduction}

In recent several years, much attention has been paid to the
challenging idea that the black hole entropy could be derived from
some symmetry inherited in general relativity {\it classically},
without knowing the details of quantum gravity. The essence of
this idea is to construct an algebra of generators associated with
a certain symmetry inherited classically in the gravity theory
considered. If such algebra obtained is a Virasoro algebra with
non-vanishing central charge, it might be true that the degeneracy
of a black hole state in the quantum theory of gravity is
determined by the central charge appeared in the algebra as in the
case of the conformal field theory through the Cardy formula. This
idea was initiated by Strominger~\cite{Strominger:1997eq} and
Birmingham et al.~\cite{BSS}. Based on the work by Brown and
Henneaux \cite{Brown:nw} that the algebra of diffeomorphisms at
spatial infinity for configurations of three dimensional
asymptotically anti-de Sitter space (AdS$_3$) induces a pair of
Virasoro algebras with nonvanishing central charge, they showed
that the application of the Cardy formula for an AdS$_3$ black
hole exactly yields the Bekenstein-Hawking entropy. Similarly, the
entropy of a cosmological horizon in de Sitter spaces has also
been reproduced in the context of de Sitter-conformal field theory
correspondence (dS/CFT correspondence)~\cite{Stro:01}. These
results are obtained also in the Chern-Simons formulation of the
three dimensional gravity with a cosmological
constant~\cite{Carlip0,BBO,MSPark}.

Unfortunately, however, these successes described above are still
incomplete for the following reasons. First of all, it does not
easily extend to black hole horizons in higher dimensional
gravitational theories. It has been shown that the algebra of
asymptotic symmetries at spatial infinity for asymptotically four
dimensional anti-de Sitter spaces is $SO(3,2)$ and does not admit
a nontrivial central extension~\cite{HT}. In addition the
Chern-Simons formulation of the general relativity in dimensions
higher than three is not known. Secondly, the conformal field
theory found in Ref.~\cite{Brown:nw} lives at the spatial infinity
while black hole entropy is expected to be related with physics on
the horizon. Recently, Carlip developed the same idea of the
algebra of diffeomorphisms, which not only is applicable to a
black hole horizon directly but also works in any higher
dimensions \cite{Carlip:1998wz,Carlip:1999cy}. He analyzed whether
the Virasoro algebra with the desirable form of a classical
central charge could arise universally on an arbitrary Killing
horizon, and also whether the Bekenstein--Hawking entropy could be
derived microscopically by the application of Cardy formula.

Lots of work in this direction have appeared
subsequently~\cite{Solo,Tries,Park:02}. However, none of them is
as yet fully satisfactory~\cite{Critics,Koga}. In particular, in
order to obtain the Virasoro algebra with the desirable form of a
central charge, which is homomorphic to Diff($S^1$) algebra up to
the central term, one has to choose one angular direction on the
horizon as in Ref.~\cite{Carlip:1999cy}. Thus, this method clearly
violates spherical symmetry in dimensions higher than three for
instance, and requires unnatural reduction of the symmetry group
on the horizon~\cite{Koga}. A framework without choosing an
angular direction is required in order to realize this idea in a
satisfactory manner.

In two dimensional spacetimes, the problem becomes more serious
since there is no room for choosing such angular direction.
Recently, by focusing on two dimensional dilaton gravity, Carlip
\cite{Carlip:2002be} suggested several new ingredients that might
lead to an improved description of the near-horizon symmetries and
possibly overcome those problems mentioned above. He claimed that,
in the presence of a black hole with a momentarily stationary
region near its horizon, the general relativity acquires a new
conformal symmetry. Moreover, a new contribution from the horizon
is added to the canonical symplectic form of general relativity so
that the central term includes an integration along the horizon.
With these new ingredients Carlip claimed that the corresponding
Virasoro algebra acquires a nonvanishing central charge and that
the Cardy formula yields the Bekenstein-Hawking entropy of a two
dimensional dilatonic black hole.

{}From the viewpoint of universality, it is of great interest to
see whether or not this new method is applicable to
higher-dimensional cases. Thus, in this paper, we explore this new
method and consider whether it works in higher-dimensional
Einstein gravity, in particular. As will be shown below
explicitly, however, the basis of functions he used to describe
such conformal transformations is singular at the horizon, and
correspondingly an integral that gives the generators of these
conformal transformations and the central charge in the Virasoro
algebra are not well-defined. Moreover, the correct use of the
eigenvalue of the generator for the zero-mode transformation
actually yields vanishing entropy. Therefore it is interesting to
check if there exists some way to resolve the problems in Carlip's
new method, including those mentioned above, with keeping the
essential features of Carlip's derivation.

In section~\ref{2D}, we briefly summarize Carlip's new derivation
for the entropy of the two-dimensional dilatonic black hole, and
point out the problems mentioned above. By choosing a different
basis which is regular even at the horizon, we show that these
problems can be resolved. In section~\ref{HigherD}, the
generalization of Carlip's derivation to any higher dimensional
pure Einstein gravity is given. In section~\ref{Dimreduction}, it
is shown that our result for the two-dimensional dilatonic black
hole entropy is consistent with that in the three-dimensional pure
Einstein gravity through dimensional reduction. Open questions and
some unsatisfactory features related to this work are discussed
finally.

\section{The two-dimensional black hole}
\label{2D}

In this section we briefly review Carlip's new approach to the
derivation of black hole entropy from symmetry for the
two-dimensional dilaton gravity~\cite{Carlip:2002be}:
\begin{equation}
I = \frac{1}{2 G} \int d^2 x \sqrt{- g} \: [\phi R +V(\phi)] .
\label{eqn:Action2DDilaton}
\end{equation}
He observes that, for field configurations in which a
``momentarily stationary'' black hole with the Killing generator
$\chi^a$ is present, conformal transformations in the form of
$\delta g_{ab} =\nabla_c(f\chi^c)g_{ab}$ together with $\delta
\phi = \nabla_c(h\chi^c)$ leave the action invariant for smooth
functions $f$ and $h$ having their support only in a small
neighborhood of the horizon. Hence he claimed that this can be
regarded as an asymptotic symmetry.\footnote{Whether or not this
is really a symmetry and so the closedness of $\mbox{\boldmath
$\omega$}$ subsequently will be discussed below.} Then, by using
the closedness of the symplectic current density $\mbox{\boldmath
$\omega$}$ associated with those transformations he also suggests
that the symplectic form of general relativity pick up a new
contribution from the horizon itself as
\begin{equation}
\hat{ \Omega} [\psi;\delta_1 \psi, \delta_2 \psi] = \int_C
\mbox{\boldmath $\omega$} [\psi; \delta_1 \psi, \delta_2 \psi ]
+\int_{\Delta} \mbox{\boldmath $\omega$} [\psi; \delta_1 \psi,
\delta_2 \psi ] .
 \label{Symplectic}
\end{equation}
Here the first integral is the standard one integrated over a
(partial) Cauchy surface $C$ and the second term is the new
contribution introduced by Carlip, which is defined as an integral
over the portion $\Delta$ of the horizon connecting a reference
cross-section $S$ and a horizon cross-section $H$ under
consideration ({\it i.e.}, an intersection of the Cauchy surface
$C$ with the horizon).

Let $l^a$ be null normal to the stretched horizon
$\tilde{\Delta}$. Then the value of the action of two-dimensional
gravity with a dilaton field $\phi$ is invariant under
transformations of
 \beqa
\delta_f g_{ab} &=& \nabla_c(fl^c)g_{ab} = (l^c\nabla_c f +kf)
g_{ab} \nonumber  \\
\delta_h \phi &=& \nabla_c (h l^c ) = l^c\nabla_c h +k h
 \label{contrans}
 \eeqa
in the asymptotic sense that the variation of the action can be
made arbitrarily small by restricting smooth functions $f$ and $h$
to have their support only in a small neighborhood $\mathcal{N}$
of the horizon, when we focus only on configurations that possess
the horizon. Here $k=\nabla_c l^c$. By rescaling the null vector
$l^a$, he takes furthermore
 \beq
\fr{k}{s} = {\rm constant\,\, on}\,\, \tilde{\Delta},
 \label{Integrability}
 \eeq
where $s \equiv l^a\nabla_a \phi \equiv \theta \phi$. Thus,
this leads that $k$ is proportional to the ``expansion'' $\theta$,
which becomes zero as the stretched horizon approaches the
horizon.

Now the variation of the generator $L[f,h]$ associated with the
transformations given by Eq. (\ref{contrans}) near the horizon is
given by~\footnote{We obtain twice of Carlip's
expression~\cite{Carlip:2002be} by straightforward calculation.}
 \beq
\delta L[f,h] = \hat{ \Omega} [\delta, \delta_{f,h}] =- \fr{1}{G}
\int_{\tilde{\Delta}} \left[ \delta\phi l^a\nabla_a (l^b\nabla_b f
+kf) -\delta k (l^b\nabla_b h +kh) \right] \hat{\bf \epsilon} ,
 \label{varHam}
 \eeq
where $\hat{\bf \epsilon}= n$ is the one-dimensional induced
volume element on $\tilde{\Delta}$ and $n^a$ is a null vector
satisfying $l^a n_a=-1$. Carlip showed that the above equation is
integrable if $\delta s$ is proportional to $\delta k$. This
integrability condition is satisfied once
Eq.~(\ref{Integrability}) holds. In addition,
Eq.~(\ref{Integrability}) also implies that
 \beq
h = \fr{s}{k} f.
 \label{Integcond}
 \eeq
The basis Carlip used is
 \beq
f_n = -\fr{\phi_+}{2\pi s} z^n , \qquad  z = e^{2\pi i \phi
/\phi_+} ,
 \label{basisCarlip}
 \eeq
where $\phi_+$ is the value of $\phi$ on the horizon \footnote{The
overall sign of $f_n$ here is opposite from Carlip's one. However,
the above choice of the sign is necessary in order to obtain the
standard sign of the first term in the right-hand side of the
Virasoro algebra Eq. (\ref{VirasoroCarlip}).}. Using
$\nabla_a\nabla_b \phi \propto g_{ab}$ on shell~\cite{Gegenetal}
and
 \beq
l\cdot\nabla z^n = \fr{2\pi in}{\phi_+} s z^n, \qquad l\cdot\nabla
s = ks
 \eeq
on the stretched horizon, we have from Eq.~(\ref{varHam})
 \beq
\delta_n L[f_m] = - \fr{2\pi i}{G} nm(n-m) \fr{s}{k}
\fr{1}{\phi_+} \int_{\tilde{\Delta}}  s z^{n+m} \,\,  \hat{\bf
\epsilon} .
 \label{varHam2}
 \eeq

Note that the integrand of this integration is regular, and
vanishes as $\tilde{\Delta} \rightarrow \Delta$. However, since
the integration limits on a portion of the stretched horizon are
chosen from $\phi =\phi_i$ on $\tilde{S}$ to $\phi =\phi_f$ on
$\tilde{H}$ with keeping $\phi_f -\phi_i =\phi_+$, the relevant
integration in Eq.~(\ref{varHam2}) becomes
 \beq
\fr{1}{\phi_+} \int_{\tilde{\Delta}}  s z^{n+m} \,\,  \hat{\bf
\epsilon} = - \fr{1}{\phi_+} \int^{\phi_f}_{\phi_i}  e^{2\pi
i(n+m)\phi /\phi_+} \,\, d\phi = - \fr{1}{2\pi i} \oint z^{n+m}
\,\, \fr{dz}{z} = - \delta_{n+m,0}
 \label{varHam3},
 \eeq
where $\hat{\bf\epsilon} = - d \phi / s$ and $\phi$ is assumed to
increase along the stretched horizon in the future. Thus we have
 \beq
\delta_{f_n} L[f_m] = \fr{4\pi i}{G} \fr{s}{k} m^3 \delta_{m+n,0}
.
 \label{Virasoro2DC}
 \eeq
If the generators form an algebra, as it is assumed by Carlip, the
central charge can be read off by comparing it with the Virasoro
algebra
 \beq
\{ L[f_m], L[f_n] \} = \delta_{f_n} L[f_m] = -i(m-n) L[f_{m+n}]
-i\fr{c}{12}m^3 \delta_{m+n,0} ,
 \label{VirasoroCarlip}
 \eeq
resulting in
 \beq
c= - \fr{48\pi}{G} \fr{s}{k} .
 \label{CC}
 \eeq
By integrating $\delta L$ in Eq.~(\ref{varHam}), Carlip also
obtained separately the eigenvalue of the operator $L$ as
 \beq
L[f_n] = -\fr{1}{G} \int_{\tilde{\Delta}} (2l \cdot \nabla s - ks)
f_n \,\, \hat{\epsilon} = \fr{1}{2\pi G} \fr{k}{s} \phi_+
\int_{\tilde{\Delta}} s z^n \,\, \hat{\epsilon} = -\fr{1}{2\pi G}
\fr{k}{s} \phi_+^2 \delta_{n,0} .
 \label{Ccharge}
 \eeq
Finally, the Cardy formula corresponding to Eq.
(\ref{VirasoroCarlip})
 \beq
\rho (\Delta) \sim \exp \left[ 2\pi \sqrt{\fr{c~\Delta}{6}}\right]
 \label{Cardy}
 \eeq
yields the entropy
 \beq
S = \mbox{log}~\rho(\Delta) =  \fr{4\pi \phi_+}{G} ,
 \eeq
where $\Delta$ is the eigenvalue of the operator $L_0 \equiv
L[f_0]$ given from Eq. (\ref{Ccharge}) as
 \beq
\Delta = - \fr{1}{2\pi G}\fr{k}{s} \phi_+^2 .
 \label{L0value}
 \eeq
This entropy is twice of the Bekenstein-Hawking entropy known for
two-dimensional dilatonic black holes.

Here we point out several flaws in Carlip's new approach briefly
summarized above. First of all, Carlip's identification of the
eigenvalue of the operator $L_0$ is somewhat erroneous. Namely,
since the $L_0$ operator is defined up to an arbitrary additive
constant in Eq. (\ref{varHam}), the direct identification of the
eigenvalue of $L_0$ operator as in Eq.~(\ref{L0value}) could be
incorrect when one apply Cardy's formula. One way to avoid such
ambiguity would be to use the result obtained in
Eq.~(\ref{Virasoro2DC}) based on the uniqueness of the central
extension of the Virasoro algebra. By comparing Eqs.
(\ref{Virasoro2DC}) with (\ref{VirasoroCarlip}), we see that the
eigenvalue $\Delta$ of the operator $L_0$ vanishes. Then, Cardy's
formula Eq. (\ref{Cardy}) indicates that the entropy actually
vanishes as well.

Secondly, notice that the base function $f_n$ in
Eq.~(\ref{basisCarlip}) diverges as the horizon is being
approached since $s \rightarrow 0$, and that $z$ is constant on
the horizon ({\it i.e.}, $z=1$) since $\phi$ is constant there. It
indicates that the integral in Eq. (\ref{varHam}) is not well
defined on the horizon. Let us consider the derivative
$l^a\nabla_a f_n$, for instance. Although the value of this
derivative cannot be computed directly on the horizon, it is clear
that this quantity is independent of $n$ since $f_n$ does not
depend on $n$ along the horizon. On the other hand, let us compute
it on the stretched horizon first and take the limit
$\tilde{\Delta} \rightarrow \Delta$. We have
 \beq
l^a\nabla_a f_n = (i n -\fr{\phi_+}{2\pi}\fr{k}{s}) z^n
\rightarrow i n -\fr{\phi_+}{2\pi} \fr{k}{s} .
 \eeq
Thus, this limiting value depends on $n$ on the horizon. Such
discrepancy implies that the derivative considered is not
continuous at the horizon and so it is ill-defined there.
Similarly, one can see that the integral in Eq.~(\ref{varHam}) is
not continuous at the horizon. Therefore, the algebra given by Eq.
(\ref{Virasoro2DC}) is actually not well-defined on the horizon.
It is not convincing at all to expect that such an ill-defined
algebra is responsible for physics on the horizon. All these
problems described above are seemingly due to the specific choices
of the rescaling of $l^a$ ({\it i.e.}, $k/s = {\rm constant}$) and
the base functions $f_n$ ({\it i.e.}, the use of a bad coordinate
$\phi$ that does not distinguish points along the event horizon).

Now we show how these flaws mentioned above can be avoided. In
Carlip's case, $k \equiv \nabla_a l^a$ vanishes at the horizon
since the rescaling freedom of the null vector $l^a$ is used to
satisfy Eq.~(\ref{Integrability}). Let us not assume such
condition for $l^a$. Instead we choose the null vector $l^a$ in
such a way that, as the horizon is being approached, it becomes
the horizon Killing generator $\chi^a$ so that the quantity $k$
becomes the surface gravity of the horizon, $\kappa$, which is a
nonvanishing constant.\footnote{The explicit form of $l^a$ in the
four-dimensional Schwarzschild black hole case, for example, is
given by Eq. (\ref{eqn:ExplicitLa}).} Defining $v$ as a non-affine
parameter describing the null trajectory on $\tilde{\Delta}$ such
that $l^a\nabla_a v =1$, we expand the transformation function $f$
in terms of mode functions given by
 \beq
f_n = -\fr{P}{2\pi} z^n , \qquad  z = e^{2\pi i v/P},
 \label{basis}
 \eeq
where  the periodicity $P$ is assumed to be an arbitrary constant
for the moment. Note that the coordinate $v$ varies along the
horizon, and these base functions are not singular at the horizon,
in contrast to Carlip's basis function. Finally we assume
 \beq
h = \alpha f,
 \eeq
where $\alpha$ is constant on $\tilde{\Delta}$. This relation
guarantees the variational equation Eq.~(\ref{varHam}) integrable,
and $\alpha =\phi_+/2$ as will be shown from dimensional reduction
below.

Here we assume that $P$ is the null distance between a reference
cross-section $\tilde{S}$ and a horizon cross-section $\tilde{H}$
measured by the function $v$ on each stretched horizon so that $z$
makes one full turn counterclockwise as $v$ runs from $\tilde{S}$
to $\tilde{H}$. Note that the null distance $P$ is taken to be
same as $\tilde{\Delta} \rightarrow \Delta$. With these
modifications we find
 \beq
\delta_{f_n} L[f_m] = -i (m-n)\fr{(P\kappa)^2}{2\pi
G}\fr{\phi_+}{2}\delta_{m+n,0} -i \fr{4\pi}{G}\fr{\phi_+}{2} m^3
\delta_{m+n,0} .
 \label{eqn:Algebra2DReg}
 \eeq
Accordingly, by comparing it with the Virasoro algebra in
Eq.(\ref{VirasoroCarlip}), one can read off
 \beq
c=\fr{24\pi \phi_+}{G}, \qquad\qquad \Delta = \fr{\pi\phi_+}{G}
\left(\fr{P\kappa}{2\pi}\right)^2 .
 \eeq
Thus the Cardy formula in Eq.(\ref{Cardy}) yields
 \beq
S =P\kappa \fr{2\pi \phi_+}{G} .
 \label{2Dentropy}
 \eeq
Notice that this entropy becomes the Bekenstein-Hawking entropy
known for the two dimensional dilaton black hole~\cite{Gegenetal}
if the periodicity could be adjusted to $P=\kappa^{-1}$. Note also
that in our method the integration of Eq.~(\ref{varHam}) gives, up
to an additive constant,
 \beq
L[f_n] = \fr{1}{G} \int_{\tilde{\Delta}} \left( 2s l\cdot \nabla
f_n +\alpha k^2 f_n \right) \, \hat{\epsilon} = \fr{(P\kappa
)^2}{2\pi G} \fr{\phi_+}{2} \,\, \delta_{n,0} .
 \eeq
Hence one finds that the eigenvalue of $L_0$ above coincides with
the one obtained from Eq.~(\ref{eqn:Algebra2DReg}).

\section{Higher-dimensional black holes}
\label{HigherD}

In this section we extend the method described in the previous
section to higher dimensional cases. Let us consider the pure
Einstein gravity with cosmological constant in an arbitrary
dimension given by
 \beq
 I = \fr{1}{16\pi G} \int d^D x \sr{-g}\,
(R-2\Lambda).  \label{action}
\eeq
 The symplectic current
($D-1$)-form for this theory may be written as
 \beq
\mbox{\boldmath $\omega$} [\psi; \delta_1 \psi, \delta_2 \psi ] =
\delta_1 {\bf \Theta} (\psi; \delta_2\psi) -\delta_2 {\bf \Theta}
(\psi; \delta_1\psi) ,
 \eeq
where $\Theta_{bcd\cdots} = \epsilon_{abcd\cdots} \Theta^a$ with
 \beq
\Theta^a = \fr{1}{16\pi G} (g_{bc} \nabla^a \delta g^{bc}
-\nabla_b \delta g^{ab}).
 \eeq
As in the previous section, we consider a symmetry variation given
by
 \beq
 \label{dg}
\delta_f g_{ab} =\nabla_c(f l^c)g_{ab}.
 \eeq
Here $l^a$ is null normal to the stretched horizon
$\tilde{\Delta}$ that becomes the horizon Killing generator
$\chi^a$ as $\tilde{\Delta} \rightarrow \Delta$. As shall be shown
below, $k \equiv \nabla^al_a$ approaches the surface gravity
$\kappa$ which is defined by $\chi^c\nabla_c \chi_b = \kappa
\chi_b$ at the horizon. In the case of the four-dimensional
Schwarzschild black hole, for instance, $l^a$ is given by
 \beq
l^a = \fr{1}{2}\left[ (\partial_t)^a +(1-r_h/r)(\partial_r)^a
\right].
 \label{eqn:ExplicitLa}
 \eeq
And at the horizon the function $v$ coincides with the ingoing
null coordinate, {\it i.e.}, $v \sim t+r_*$ where $r_* =r_h \ln
(r-r_h)$ is the usual ``tortoise'' coordinate. The variation of
the generator $L[f]$ is now given as
 \beqa
\delta L[f] &=& -\fr{(D-1)(D-2)}{32\pi G~D } \int_{\tilde{\Delta}}
\left[ g^{ab}\delta g_{ab}\,\, l^c \nabla_c (l^d \nabla_d f +kf)
-l^c \nabla_c (g^{ab}\delta g_{ab})\,\, (l^d\nabla_d f +kf)
\right]
\hat{\bf \epsilon}   \nonumber \\
&=& -\fr{(D-1)(D-2)}{16\pi G~D } \int_{\tilde{\Delta}}
g^{ab}\delta g_{ab}\,\, l^c \nabla_c (l^d \nabla_d f +kf) \hat{\bf
\epsilon},
 \label{generator}
 \eeqa
where we have used the integration by parts and $\hat{\bf
\epsilon}$ is the ($D-1$)-dimensional induced volume element on
$\tilde{\Delta}$. One can check that the variational form of the
generator $L$ in Eq. (\ref{generator}) is actually integrable,
when we focus on a narrow subspace of the phase space and assume
that all relevant variations are described by Eq. (\ref{dg}).
Namely, when the variation $\delta$ is thought of a derivative on
the space of metric fields, an explicit calculation shows
$\delta_1 (\delta_2L)-\delta_2(\delta_1L)=0$. Thus, one does not
require any further condition for the integrability.

As in Eq.~(\ref{basis}), we choose a basis of functions as
 \beq
f_n = -\fr{P}{\pi D} z^n , \qquad  z = e^{2\pi i v/P} ,
 \label{basisD}
 \eeq
where the normalization is chosen such that the base function
$f_n$ satisfies the commutation relations isomorphic to the
Diff($S^1$) algebra,
 \beq
\{ f_m, f_n \} = i (n-m) f_{m+n},
 \label{DiffS1}
 \eeq
with the brackets between the basis functions defined through
 \beq
[ \delta_{f_m} , \delta_{f_n} ] g_{a b} \equiv \delta_{\{ f_m , f_n \}} g_{a b} .
 \eeq
Now one can explicitly obtain from Eq. (\ref{generator}) that
 \beqa
\delta_{f_m} L[f_n] &=& -\fr{(D-1)(D-2)}{16\pi G}
\int_{\tilde{\Delta}} (l^c\nabla_c f_m +kf_m) \,\, l^c \nabla_c
(l^d \nabla_d f_n +kf_n) \hat{\bf
\epsilon} \nonumber  \\
&=&- \fr{(D-1)(D-2)}{ D^2}\fr{1}{ 4\pi G} \int_{\tilde \Delta}
\Bigg[ -mn^2 + n
\left(\fr{Pk}{2\pi}\right)^2 + \fr{Pki}{2\pi} n(m+n)  \nonumber \\
 & & \quad\qquad\qquad\qquad\qquad +\fr{P^2
}{4\pi^2} \left(m-\fr{Pki}{2\pi}\right) l^c\nabla_c k \Bigg]
z^{m+n-1} dz~ d\Sigma .
 \label{Virasoro}
 \eeqa
Here $d\Sigma$ is the infinitesimal volume element of the spatial
cross section of the stretched horizon $\tilde{\Delta}$, and
$dv=Pdz/2\pi i z$.

The integrand above is not constant in general. As the stretched
horizon approaches the horizon ({\it i.e.}, $\tilde{\Delta}
\rightarrow \Delta$), however, one can see that the quantity
$k=\nabla^cl_c$ becomes the surface gravity $\kappa$. Let the null
vector $n^a$ be tangent to the ingoing null trajectory and be
scaled such that $n^cl_c=-1$, and let $\sigma_{a b}$ be the
induced metric on the spatial cross section of the stretched
horizon as $\sigma_{a b} \equiv g_{a b} + l_a n_b + l_b n_a$. Then
 \beqa
k = g^{ab} \nabla_a l_b &=& ( \sigma^{a b} -l^an^b -l^bn^a)
\nabla_a l_b
\nonumber  \\
&=& \theta -n^bl^a\nabla_a l_b  \nonumber \\
&\rightarrow & \kappa .
 \eeqa
Here we used that $l^bn^a \nabla_a l_b =n^a\nabla_a (l^bl_b)/2=0$,
that the expansion of the congruence of null geodesics $\theta =
\sigma^{a b}\nabla_a k_b = \sigma^{a b}\nabla_al_b$, where $k^a$
is the geodesic tangent, vanishes at the horizon, and that $l^a
\rightarrow \chi^a$ and $l^a \nabla_a l_b \rightarrow \chi^a
\nabla_a \chi_b$ as $\tilde{\Delta} \rightarrow \Delta$. The
surface gravity of a stationary event horizon may be defined as
$\chi^c\nabla_c \chi_b = \kappa \chi_b$, which is constant along
the horizon provided that the dominant energy condition is
satisfied. Thus, the integrand becomes constant as the stretched
horizon approaches the event horizon. Finally, from the same
calculation as in the previous section, we have at the horizon
 \beqa
\delta_{f_n} L[f_m] &=&
\fr{(D-1)(D-2)}{D^2} \fr{A}{4G}   \nonumber  \\
& & \times \left[ -i(m-n) \left(\fr{P\kappa}{2\pi}\right)^2
\delta_{m+n,0} -2i m^3 \delta_{m+n,0} \right],
 \label{VirasoroD}
 \eeqa
where $A=\oint d\Sigma$ is the surface area of the
$(D-2)$-dimensional horizon cross section.

By comparing it with Eq.~(\ref{VirasoroCarlip}), therefore, we
obtain the nonvanishing central charge given by
 \beq
c =\fr{24(D-1)(D-2)}{D^2} \fr{A}{4G},
 \label{central}
 \eeq
and the eigenvalue of the $L_0$ operator given by
 \beq
\Delta = \left(\fr{P\kappa}{2\pi}\right)^2 \fr{(D-1)(D-2)}{D^2}
\fr{A}{4G} = \left(\fr{P\kappa}{2\pi}\right)^2 \fr{c}{24} .
 \label{Lzero}
 \eeq
By applying the Cardy formula for the density of states in
Eq.~(\ref{Cardy}), the entropy becomes
 \beq
S = \mbox{log} ~\rho (\Delta)= \fr{2(D-1)(D-2)P\kappa}{D^2}
\fr{A}{4G} .
 \label{BHE}
 \eeq
This may be adjusted to the Bekenstein-Hawking entropy if the
periodicity could be chosen such that
 \beq
P= \fr{D^2}{2(D-1)(D-2)} \kappa^{-1} .
 \label{Period}
 \eeq

\section{Dimensional reduction}
\label{Dimreduction}

Since the two-dimensional dilaton gravity can be obtained from a
dimensional reduction of a higher-dimensional pure Einstein
gravity, it is interesting to see whether our results for the
entropies obtained above are consistent in this context. If we
consider three-dimensional black holes for simplicity, the entropy
is given by
\begin{equation}
S = \frac{4}{9} \: P\kappa \: \frac{A}{4 G}
 \label{eqn:Entropy3D}
\end{equation}
from Eq.~(\ref{BHE}). We ``$2 + 1$'' decompose the
three-dimensional metric $g_{a b}$ as
 \beq
 \label{ds3}
g_{a b} = h_{a b} + m_{a} m_{b},
 \eeq
where the unit normal $m_{a}$ to 2D-subspace is given as
 \beq
m_a = r \nabla_a \varphi ,
 \label{eqn:normalDR}
 \eeq
in terms of the radius $r$ and the angular coordinate $\varphi$ of
the circles in the third dimension, and $h_{a b}$ is the induced
metric of the two-dimensional subspace. Since $r$ plays the role
of the ``lapse'' function for the ``evolution'' in the
$\varphi$-direction, we can write as
 \beq
\sqrt{- g} \; {}^{(3)}\!R = \sqrt{- h} \: r \: \left( {}^{(2)}\!R
+ \mathcal{L} \right),
 \eeq
where $\mathcal{L} = K_{ab}K^{ab}-K^2$ denotes terms consisting of
the extrinsic curvature of a $\varphi = {\rm constant}$ surface.
We rewrite the radius $r$ in terms of the dilaton field $\phi$ as
$r = \phi$. Then, the terms of the extrinsic curvature can be
considered as matter parts. As has been shown in the N\"{o}ther
charge method of black hole entropy \cite{WI,JKM}, this matter
action does not change the entropy result. Accordingly we ignore
the terms of the extrinsic curvature and see if this is consistent
with the entropies obtained in the previous sections.

By assuming configurations independent of $\varphi$, we have
\begin{eqnarray}
\label{eqn:Action3D}
I & = & \frac{1}{16 \pi G} \int \, d^3 x
\sqrt{- g} \; ({}^{(3)}\!R -2 \Lambda)  \\
& = & \frac{1}{16 \pi G} \int \, d\varphi \, d^2 x \sqrt{- h} \:
\phi \: ({}^{(2)}\!R -2 \Lambda +\mathcal{L}[\phi]
) \nonumber \\
& = & \frac{1}{8 G} \int \, d^2 x \sqrt{- h} \: \phi \:
({}^{(2)}\!R -2 \Lambda ) .
\label{eqn:ActionReduced}
\end{eqnarray}
Notice that Eq.(\ref{eqn:ActionReduced}) is a quarter of
Eq.(\ref{eqn:Action2DDilaton}), and so are the values of $c$ and
$\Delta$. Then, from the two-dimensional result
Eq.~(\ref{2Dentropy}) the entropy is given by
\begin{equation}
S = P \kappa \frac{\pi \phi_+}{2G},
 \label{2DBHE}
\end{equation}
if $l^a$ is null and approaches the Killing vector also in the
two-dimensional subspace. Now we expect that this should be
equivalent to the entropy of the three dimensional black hole. By
substituting $\phi_+ = r_+$, Eq.(\ref{2DBHE}) becomes
 \beq
S = P\kappa \frac{\pi r_+}{2 G}
 =  P\kappa \frac{A}{4 G}  ,
 \label{eqn:EntropyReduced}
 \eeq
where $A = 2 \pi r_+$ is the ``area'' (i.e., circumference) of the
horizon in the three dimension. One can see that this coincides
with the full three dimensional result in Eq.(\ref{eqn:Entropy3D})
since the difference by a numerical factor $4/9$ simply comes from
the dimension dependent normalization of the mode function $f_n$
in Eq.~(\ref{basisD}). Therefore, our method gives the consistent
result under dimensional reduction. Since the additional matter
action resulting from the dimensionally reduced theory is ignored,
our result also indicates that the matter action does not
contribute to the entropy as it happens in the usual cases.

Carlip~\cite{Carlip:2002be} imposed the relationship between $f$
and $h$ given by Eq.(\ref{Integcond}) in order to make the
conformal transformations integrable. In the context of
dimensional reduction explained above, it is expected that the
variation of the dilaton field $\phi$ in the two-dimensional
theory can be deduced from that of the metric field in the
three-dimensional theory. By requiring the normalization condition
$m^a m^b g_{a b} = 1$ is preserved under the variations, we first
find from Eq. (\ref{dg}) and the variation of Eq.
(\ref{eqn:normalDR}) that
 \beq
\nabla_c ( f l^c ) = 2 \frac{\delta_h \phi}{\phi} ,
 \eeq
where the variation induced on $\phi$ is denoted as $\delta_h
\phi$, and $r = \phi$ as well as a relation between $f$ and $h$,
which is to be determined, is understood. On the other hand, from
the fact that $l^a$ is null in both three-dimensions and
two-dimensions, we see that $l^a$ should reside in the
two-dimensional subspace as $m_a l^a = 0$, and then we can show by
using it that
 \beq
\nabla_c (f l^c ) = \frac{1}{\phi} D_c ( \phi f l^c ) ,
 \eeq
where $D_c$ is the covariant derivative associated with the
induced metric $h_{a b}$. Since $\delta_h \phi$ given by Eq.
(\ref{contrans}) is written as $\delta_h \phi = D_c ( h l^c )$, we
thus find
 \beq
h= \fr{\phi}{2} f,
 \eeq
which is consistent with the integrability condition in the
two-dimensional case.\,\footnote{With this relationship one can
see that $\delta s = \phi \delta \kappa /2 + s \nabla_c(fl^c)$.
Thus, $\delta s$ becomes proportional to $\delta \kappa$ as the
horizon is being approached since $s \rightarrow 0$ and
$\nabla_c(fl^c)$ is regular.}

\section{Discussion}

To conclude, we have analyzed whether the entropy of black holes
for the pure Einstein gravity in any dimensions can be derived
from a Virasoro algebra associated with a specific class of
near-horizon conformal transformations given in Eq.~(\ref{dg}). We
simply extended Carlip's derivation developed for two-dimensional
dilaton black holes in Ref.~\cite{Carlip:2002be}. However, there
are some important modifications in choices of the null vector
$l^a$ and the base function $f_n$ as in
Eqs.~(\ref{eqn:ExplicitLa}) and (\ref{basisD}). As can be seen in
Eq.~(\ref{BHE}), the entropy obtained is proportional to the area
of the event horizon, but it also depends linearly on the product
of the surface gravity and the parameter length of a horizon
segment in consideration ({\it i.e.}, $\sim P\kappa$).

The entropy derivation explained above does not depend on the
details of black hole solutions. What we actually need is simply a
neighborhood of a Killing horizon. Namely the horizon is a null
hypersurface which is generated by a Killing vector field.
Therefore it is straightforward to apply the same method to other
types of horizons such as a Rindler horizon or a de Sitter
horizon. Recently, a generalization of the black hole
thermodynamics to any ``causal horizon'' has been argued in
Ref.~\cite{Jacobson}.

Another feature of this derivation is that, as can be seen in
Eqs.~(\ref{generator}) and (\ref{Virasoro}), the central charge is
given by
 \beq
K[f_1,f_2] \sim \int dv\, d\Sigma \,(D\!f_1\, D^2\!f_2 -D\!f_2\,
D^2\!f_1) ,
 \eeq
where $D =l\cdot\nabla =\partial_v$. This becomes the standard
form of the central term for 2D conformal field theory. Note that
it contains an integration over the ``time'' parameter $v$
explicitly in contrast to other approaches as in
Refs.~\cite{Carlip:1998wz,Carlip:1999cy}. In particular, in
Ref.~\cite{Carlip:1999cy} an orthogonality condition for base
functions was used to recover a conventional Virasoro algebra.
Although fixing the average surface gravity on the horizon
cross-section can naturally lead to such an orthogonality
relation, physical origin of such a boundary condition still
remains unclear. The orthogonality relations in
Refs.~\cite{Carlip:1998wz,Carlip:1999cy} were realized by adding
extra angular dependence to the $r$-$t$ diffeomorphisms. As
mentioned above, however, such prescription not only requires an
unnatural reduction of the symmetry group, but also cannot work
for two-dimensional cases. The new entropy derivation described in
the present work is free from the problem of adding extra angular
dependence. This new feature is essentially due to that the newly
added contribution to the symplectic form in
Eq.~(\ref{Symplectic}) has an integration along the horizon from
the beginning.

However, it is important to note that the conformal
transformations we considered cannot be described by
diffeomorphisms. One way to see this is the following. Suppose
that the transformations are indeed described by diffeomorphisms.
Then the integrability condition derived by Wald and
Zoupas~\cite{Wald:1999wa} for the generator $L$ should be
satisfied. The condition is
 \beq
\int_{\partial \Sigma} \xi\! \cdot \mbox{\boldmath $\omega$}=0 ,
 \label{intcon}
 \eeq
where $\xi$ is a vector field generating diffeomorphisms and
$\Sigma$ a Cauchy surface. Since the part defined over the horizon
$\Delta$ in Eq.~(\ref{Symplectic}) was shown to be integrable
separately, this condition becomes equivalent to $\int_{\hat{H}}
\xi\! \cdot \mbox{\boldmath $\omega$}=0$ where $\hat{H}$ is a
horizon cross-section which is the inner boundary of $C$. Hence
the contribution to the variation of the generator $L$ arising
from the integral along the null direction on the horizon should
vanish since
 \beq
\delta L_{\xi} = \int_{\Delta} \mbox{\boldmath $\omega$} \sim
\int^{H}_{S} dv \int_{\hat{H}} \xi\! \cdot \mbox{\boldmath
$\omega$} .
 \eeq
However the explicit calculation of the variation of $L$ shows
that it does not vanish as in Eq.~(\ref{VirasoroD}). Thus the
conformal transformations considered in this paper cannot be
regarded as diffeomorphisms such as ``conformal isometries''
generated by conformal Killing vectors in the vicinity of the
horizon. Therefore, the algebra appeared in the present
formulation is quite different from those associated with
diffeomorphisms in the literature so far.

Since the near-horizon conformal transformations considered in
this paper are not diffeomorphisms as argued above, the standard
covariant phase space method \cite{WI,Lee:90} that we have adopted
in this paper might not work. For example, the linearized field
equations for $\delta g_{ab}$ are not ensured to hold in general.
Consequently the symplectic current $\mbox{\boldmath $\omega$}$ is
not necessarily closed. The new idea to add the integral along the
null direction to the variation of $L$ as in
Eq.~(\ref{Symplectic}) might then be unjustifiable. As a related
issue, we should point out that there does not exist any
convincing reasonings to consider the transformations proposed by
Carlip and us are N\"{o}ther's symmetry transformations. Although
N\"{o}ther's symmetry transformations make the action invariant
for arbitrary background configurations, i.e., N\"{o}ther's
symmetry is a symmetry of the action itself without depending on
configurations, neither Carlip's nor our transformations possess
this property. (See Ref.~\cite{Carlip:2002be} for Carlip's
argument, where ``invariance of action'' is shown only for
configurations that have a special property, not for arbitrary
configurations.) Then, the generators of these transformations are
not ensured to be closed, and hence not ensured to form an algebra
as we assumed. Therefore, we should explore further the reasonings
to justify the method proposed by Carlip and extended by us in
this paper, as Carlip himself mentioned in
Ref.~\cite{Carlip:2002be}. In addition, when we show the
integrability of the operator $L$, we assumed that all variations
are described by the conformal transformations given by
Eq.~(\ref{dg}). Thus, the subspace of the phase space we
considered is quite narrow, compared to the covariant phase space
in Ref.~\cite{Lee:90}. It should be clarified whether
quantization, or defining eigenstates at least, in this narrow
subspace can be carried out in a manner similar to the standard
Dirac method.

It should be pointed out that, in addition to the usual
Bekenstein-Hawking entropy, the entropy result Eq.~(\ref{BHE})
obtained in our method has a multiplication factor that is a
function of the periodicity $P$, the surface gravity $\kappa$, and
the spacetime dimension $D$. Recall that $P$ is the parameter
length between a reference horizon cross-section $S$ and the
horizon cross-section $H$ in consideration. Since the horizon
cross-section can be chosen arbitrarily on the horizon, such
explicit $P$-dependence in the entropy derivation is highly
unsatisfactory. In Carlip's method~\cite{Carlip:2002be}, the
parameter length between any two points on the horizon vanishes
when measured in $\phi$, since $\phi$ is constant on the horizon.
Then, in order to keep $P=\phi_f -\phi_i =\phi_+$ fixed in
Eq.~(\ref{varHam3}) the position of the reference horizon
cross-section must move towards the past as the limiting process
$\tilde{\Delta} \rightarrow \Delta$ is taken. In other words, in
Carlip's method~\cite{Carlip:2002be} it is likely that the
reference horizon cross-section should locate somewhere in the
past ``infinity'' (or in the future ``infinity'') on the horizon,
while a ``momentarily stationary'' region, i.e., a finite segment
of the horizon, is considered. Note that smoothness of the
Euclidean sector of near-horizon geometry naturally requires the
periodicity of $\sim \kappa^{-1}$. Thus, one might expect that it
will be possible to choose a specific value of the periodicity $P$
such that the multiplication factor becomes unit as in
Eq.~(\ref{Period}). It implies that for a given black hole the
Bekenstein-Hawking entropy is reproduced in our derivation only
when a specific length of the horizon segment, which depends on
the surface gravity and the spacetime dimension through
Eq.~(\ref{Period}), is taken. There is no clear physical
explanation at the moment for why such specific value should be
chosen for the periodicity though.

We should mention that the method presented in this paper yields
vanishing entropy in 2D pure Einstein gravity, as we can see from
Eq.~(\ref{BHE}). Since the field equation is trivially satisfied
in this case, it may indicate that thermodynamic entropy in this
theory is not well-defined. Then, it might make no sense to
consider horizon entropy in 2D pure Einstein gravity. On the other
hand, however, a horizon is shown to be a thermal object even in
2D Einstein gravity without using the field equation of gravity,
and hence vanishing statistical entropy looks inconsistent with
this fact since it implies a frozen object. Actually, we
considered a sort of statistical entropy in this paper, which has
not been shown to coincide with the thermodynamic entropy. The
ill-defined classical dynamics may be simply a problem of the
thermodynamic entropy, not of the statistical entropy. From this
point of view, vanishing entropy result in the 2D pure Einstein
gravity seems to signal failure of the method proposed by Carlip
and extended in this paper.

{}From the physical point of view, it is also of interest to
consider what ``microscopic states'' are responsible for black
hole entropy. We see that the zero-mode of the conformal
transformations, which is generated by $L_0$,  rescales the metric
by a constant. Since the entropy derived above is defined as the
logarithm of the number of the eigenstates of $L_0$, this
indicates that the ``microscopic states'' responsible for black
hole entropy are the eigenstates of ``rescaling of the metric by a
constant''. It does not seem to have something to do with
properties of a black hole horizon, such as energy and angular
momentum.

Therefore, the method analyzed in this paper possesses some
unsatisfactory features, and further investigation is necessary in
the future to understand whether this method is really correct or
not.

\section*{Acknowledgments}

GK would like to thank Seungjoon Hyun, Makoto Natsuume and Don
Page for useful discussions. JK also would like to thank Makoto
Natsuume for helpful discussions and warm hospitality at KEK. GK
was in part supported by JSPS (Japanese Society for Promotion of
Sciences) during the early stage of this work. MIP was supported
by the Korea Research Foundation Grant (KRF-2002-070-C00022).

\newcommand{\J}[4]{#1 {\bf #2} #3 (#4)}
\newcommand{\andJ}[3]{{\bf #1} (#2) #3}
\newcommand{\AP}{Ann.\ Phys.\ (N.Y.)}
\newcommand{\MPL}{Mod.\ Phys.\ Lett.}
\newcommand{\NP}{Nucl.\ Phys.}
\newcommand{\PL}{Phys.\ Lett.}
\newcommand{\PR}{Phys.\ Rev.\ D}
\newcommand{\PRL}{Phys.\ Rev.\ Lett.}
\newcommand{\PTP}{Prog.\ Theor.\ Phys.}
\newcommand{\hep}[1]{ hep-th/{#1}}
\newcommand{\hepp}[1]{ hep-ph/{#1}}
\newcommand{\hepg}[1]{ gr-qc/{#1}}

\end{document}